\documentclass[aps,prb,twocolumn,superscriptaddress,showpacs,amsmath,amssymb,floatfix]{revtex4}

\usepackage{graphicx}

\newcommand{\paren}[1]{\ensuremath{\left( #1 \right)}}
\newcommand{\brac}[1]{\ensuremath{\left[ #1 \right]}}
\newcommand{\abs}[1]{\ensuremath{\left| #1 \right|}}
\newcommand{\bra}[1]{\ensuremath{\left< #1 \right|}}
\newcommand{\ket}[1]{\ensuremath{\left| #1 \right>}}
\newcommand{\partiald}[2]{\ensuremath{\frac{\partial #1}{\partial #2}}}

\newcommand{\matel}[3]{\ensuremath{\left< #1 \left| \, #2 \, \right|
                                    #3 \right>}}
\newcommand{\trans}[2]{\ensuremath{#1 \rightarrow #2}}

\newcommand{\eq}[1]{(\ref{eq#1})}
\newcommand{\eqa}[1]{Eq.~\eq{#1}}

\newcommand{\Ham}{\ensuremath{\mathcal{H}}}

\newcommand{\Gn}[1]{\ensuremath{\Gamma_{#1}}}
\newcommand{\wnm}[2]{\ensuremath{\omega_{#1 #2}}}
\newcommand{\dm}{\ensuremath{\rho}}
\newcommand{\wrf}{\ensuremath{\omega_{r\!f}}}
\newcommand{\Irf}{\ensuremath{I_{r\!f}}}
\newcommand{\Crf}{\ensuremath{C_{r\!f}}}
\newcommand{\Phio}{\ensuremath{\Phi_0}}

\newcommand{\Geq}{\ensuremath{\Gamma_{\infty}}}

\newcommand{\Rnm}[2]{\ensuremath{\Omega_{#1 #2}}}
\newcommand{\RRnm}[2]{\ensuremath{\Omega^{min}_{R,#1 #2}}}

\newcommand{\RReffnm}[2]{\ensuremath{\Omega_{R,#1 #2}}}

\newcommand{\Ps}{\ensuremath{P_S}}

\newcommand{\rmunit}[1]{\ensuremath{\mathrm{#1}}}
\newcommand{\um}{\ensuremath{\mu \rmunit{m}}}
\newcommand{\uA}{\ensuremath{\mu \rmunit{A}}}
\newcommand{\nA}{\ensuremath{\rmunit{nA}}}

\newcommand{\ns}{\ensuremath{\rmunit{ns}}}
\newcommand{\nH}{\ensuremath{\rmunit{nH}}}
\newcommand{\pH}{\ensuremath{\rmunit{pH}}}
\newcommand{\fF}{\ensuremath{\rmunit{fF}}}
\newcommand{\pF}{\ensuremath{\rmunit{pF}}}
\newcommand{\Acm}{\ensuremath{\rmunit{A} / \rmunit{cm}^2}}
\newcommand{\GHz}{\ensuremath{\rmunit{GHz}}}
\newcommand{\MHz}{\ensuremath{\rmunit{MHz}}}
\newcommand{\dBm}{\ensuremath{\rmunit{dBm}}}
\newcommand{\invs}{\ensuremath{\rmunit{1/s}}}
\newcommand{\mW}{\ensuremath{\rmunit{mW}}}
\newcommand{\AW}{\ensuremath{\nA / \sqrt{\mW}}}
\newcommand{\mK}{\ensuremath{\rmunit{mK}}}
\newcommand{\As}{\ensuremath{\rmunit{A} / \rmunit{s}}}

\newcommand{\UMCP}{Center for Nanophysics and Advanced Materials and Joint Quantum Institute, Department~of~Physics, University of Maryland, College Park, Maryland 20742-4111, USA}
\newcommand{\NIST}{National Institute of Standards and Technology, Gaithersburg, Maryland 20899-8423, USA}

\begin{document}

\title{Multilevel effects in the Rabi oscillations of a Josephson phase qubit}

\author{S.~K.~Dutta}
\affiliation{\UMCP}
\author{Frederick~W.~Strauch}
\affiliation{\NIST}
\author{R.~M.~Lewis}
\affiliation{\UMCP}
\author{Kaushik~Mitra}
\affiliation{\UMCP}
\affiliation{\NIST}
\author{Hanhee~Paik}
\affiliation{\UMCP}
\author{T.~A.~Palomaki}
\affiliation{\UMCP}
\author{Eite~Tiesinga}
\affiliation{\UMCP}
\affiliation{\NIST}
\author{J.~R.~Anderson}
\author{Alex~J.~Dragt}
\author{C.~J.~Lobb}
\author{F.~C.~Wellstood}
\affiliation{\UMCP}

\date{\today}

\begin{abstract}
We present Rabi oscillation measurements of a Nb/AlO$_\text{x}$/Nb dc superconducting quantum interference device (SQUID) phase qubit with a 100 $\um^2$ area junction acquired over a range of microwave drive power and frequency detuning.  Given the slightly anharmonic level structure of the device, several excited states play an important role in the qubit dynamics, particularly at high power.  To investigate the effects of these levels, multiphoton Rabi oscillations were monitored by measuring the tunneling escape rate of the device to the voltage state, which is particularly sensitive to excited state population.  We compare the observed oscillation frequencies with a simplified model constructed from the full phase qubit Hamiltonian and also compare time-dependent escape rate measurements with a more complete density-matrix simulation.  Good quantitative agreement is found between the data and simulations, allowing us to identify a shift in resonance (analogous to the ac Stark effect), a suppression of the Rabi frequency, and leakage to the higher excited states.
\end{abstract}

\pacs{74.50.+r, 03.67.Lx, 85.25.Dq}

\maketitle


\section{Introduction}
\label{SIntro}

Over the past decade, a variety of qubits based on superconducting Josephson junctions has been proposed and experimentally realized.\cite{Makhlin01a, Devoret04a, You05a, Wendin07a, Zagoskin07a}  Among them is the phase qubit, which can take the form of a single current-biased junction,\cite{Ramos01a, Yu02a} a flux-biased rf superconducting quantum interference device (SQUID),\cite{Simmonds04a} a low inductance dc SQUID,\cite{Claudon04a} a large inductance dc SQUID where one of the junctions is thought of as the qubit,\cite{Martinis02a} or a vortex in an annular junction.\cite{Wallraff03c}  The behaviors of the devices are similar, as they all employ a large area junction, where the dynamics of the quantum mechanical phase difference across the junction (or the orientation of the vortex in the last case) are determined by a tilted washboard-like potential.  The two lowest states in a well of this potential serve as the qubit basis.  Among the possible benefits of the phase qubit are its relative insensitivity to charge and flux noise and its ability to operate over a wide range of parameters.  Recent demonstrations include fast readout,\cite{Claudon04a} Rabi oscillations,\cite{Yu02a, Martinis02a, Claudon04a, Strauch06a, Lisenfeld07a} and simple capacitive coupling of two qubits as measured through spectroscopy,\cite{Xu05a} simultaneous state measurement,\cite{McDermott05a} and state tomography.\cite{Steffen06b}

In the phase qubit, higher excited states can impact the dynamics strongly, due to the nearly harmonic level structure within the potential well.  The presence of these quantized energy levels has been clearly detected by monitoring the decay of a thermal population\cite{Silvestrini97a} and by microwave spectroscopy with single\cite{Martinis85a} and multiphoton\cite{Wallraff03a, Strauch06a} transitions.  Multilevel systems can, for example, be exploited for state initialization\cite{Valenzuela06a} and readout,\cite{Martinis02a} quantum logic gates\cite{Amin03a, Strauch03a, Wei08a} and algorithms,\cite{Ahn00a} and cryptography.\cite{Groblacher06a}  However, when controlling the state of the phase qubit with a microwave current, off-resonant excitation of higher levels leads to leakage out of the desired qubit space and therefore a loss of quantum information.

Rabi oscillations serve as both a standard demonstration of quantum state manipulation and a diagnostic for decoherence and control fidelity.  While these oscillations take on a simple form in a two-level system, several effects can occur in a multilevel system: the oscillations can distort, the higher states can become populated, and the basic resonance properties of the oscillations can undergo subtle shifts.  Ultimately, to achieve fast and accurate control of the qubit state, all of these effects need to be carefully characterized.  Progress in this area has been predominantly theoretical in nature, pointing out the conditions under which errors are introduced and methods to minimize their effects.\cite{Goorden03a, Steffen03a, Meier05a, Amin06a, Strauch06a, Shevchenko07a}  There has, however, been less direct experimental evidence confirming the validity of models upon which these predictions are based,\cite{Claudon04a, Claudon07a, Strauch06a, Lucero08a} particularly with regard to leakage.

In this paper, we investigate the influence of the higher states of a phase qubit by examining the behavior of Rabi oscillations at high power, where their impact is greatest.  In Sec.~\ref{SPhase}, we describe the design of our qubit and a Hamiltonian that approximates its dynamics.  Also discussed are various experimental details and the readout scheme that allows measurement of very small excited state populations.  Section~\ref{SStrong} describes Rabi oscillations taken at a range of power and detuning from resonance, along with comparisons to a simple theory, while in Sec.~\ref{SDecoherence} we develop a more complete density-matrix model including the effects of decoherence and noise.  Finally, Sec.~\ref{SSummary} contains a summary of the key points of our findings.

\section{\lowercase{dc} SQUID Phase Qubit}
\label{SPhase}

\begin{figure}
   \includegraphics[bb=0 0 335 571, clip, width=3.0in]{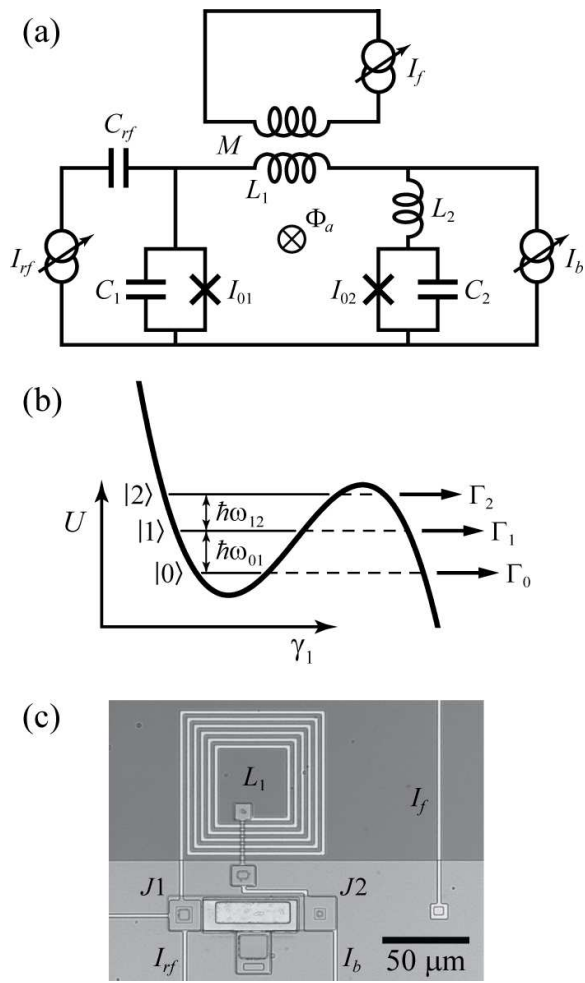} \vspace{-0.in}
  \caption{\label{FDevice}The dc SQUID phase qubit.  (a) The qubit junction $J1$ (with critical current $I_{01}$ and capacitance $C_1$) is isolated from the current bias leads by an auxiliary junction $J2$ (with $I_{02}$ and $C_2$) and geometrical inductances $L_1$ and $L_2$.  The device is controlled with a current bias $I_b$ and a flux current $I_f$ which generates flux $\Phi_a$ through mutual inductance $M$.  Transitions can be induced by a microwave current \Irf, which is coupled to $J1$ via \Crf.  (b) When biased appropriately, the dynamics of the phase difference $\gamma_1$ across the qubit junction are analogous to those of a ball in a one-dimensional tilted washboard potential $U$.  The metastable state \ket{n}\ differs in energy from \ket{m}\ by $\hbar \wnm{n}{m}$ and tunnels to the voltage state with a rate \Gn{n}.  (c) The photograph shows a Nb/AlO$_\text{x}$/Nb device.  Not seen is an identical SQUID coupled to this device intended for two-qubit experiments; the second SQUID was kept unbiased throughout the course of this work.}
\end{figure}

Figure~\ref{FDevice}(a) shows the circuit schematic for our dc SQUID phase qubit.\cite{Palomaki06a}  The qubit junction $J1$ (with critical current $I_{01}$ and capacitance $C_1$) is shown on the left.  It is isolated from the current bias source $I_b$ by geometrical inductances $L_1$ and $L_2$ and the second junction $J2$ (with $I_{02}$ and $C_2$).  In order to independently control the currents in the two arms of the resulting dc SQUID, a current source $I_f$ applies a flux $\Phi_a$ to the SQUID loop through mutual inductance $M$.  Good isolation of the qubit junction is obtained when $L_1 / M \gg 1$ and $L_1 / \paren{L_2 + L_{J2}} \gg 1$, where $L_{J2}$ is the Josephson inductance of the isolation junction.\cite{Martinis02a}

For arbitrary values of the bias current $I_b$ and flux current $I_f$, the dynamics of a dc SQUID can be described by 2 degrees of freedom corresponding to the phase differences across each of the junctions.  We, however, operate the device by increasing $I_b$ by $\Delta I_b$ while simultaneously increasing $I_f$ by $L_1 \Delta I_b / M$.  This nominally keeps the total current through the isolation junction $J2$ near zero, so that the qubit junction current is roughly $I_b$.  Furthermore, one obtains a weak dynamical coupling between the junctions by choosing $L_1$ to be large and biasing the SQUID so that the two junctions are well out of resonance with each other.\cite{Palomaki06a}  In this case, the dynamics of the phase difference $\gamma_1$ across the qubit junction are governed to a good approximation\cite{Mitra06a} by the Hamiltonian of a single current-biased junction,\cite{Fulton74a, Leggett87a}
\begin{equation}
  \Ham = \frac{4 E_C}{\hbar^2} p_1^2 - E_J \paren{\cos \gamma_1
         + \frac{I_b - \Irf \cos \wrf t}{I_{01}} \gamma_1}.
\label{eqa}
\end{equation}
Here, $E_C = e^2 / 2 C_1$ and $E_J = I_{01} \Phio / 2 \pi$ are the charging energy and Josephson coupling energy of the qubit junction, $p_1 = \paren{\Phio / 2 \pi}^2 C_1 \dot{\gamma_1}$ is the momentum conjugate to $\gamma_1$, and \Irf\ is the amplitude of a microwave drive current of frequency \wrf.  In quantizing $\Ham$, $\gamma_1$ and $p_1$ become operators with $\brac{\gamma_1, p_1} = i \hbar$.

The second term on the right-hand side of \eqa{a}\ defines a one-dimensional tilted washboard potential $U$, sketched in Fig.~\ref{FDevice}(b).  Each well is characterized by the classical plasma frequency $\omega_p$ and barrier height $\Delta U$, which can also be expressed as the dimensionless quantity $N_s = \Delta U / \hbar \omega_p$.  A single potential well supports roughly $N_s$ metastable states \ket{n}, where the ground state \ket{0}\ and first excited state \ket{1}\ serve as the basis for quantum computation.\cite{Ramos01a}  Motivated by this, the Hamiltonian can be expressed in a discrete representation as
\begin{equation}
  \Ham_N = \sum_{n=0}^{N-1} \hbar \wnm{0}{n} \ket{n} \bra{n}
         + \sum_{n, m = 0}^{N-1} \hbar \Rnm{n}{m} \ket{n} \bra{m} \cos \wrf t,
\label{eqc}
\end{equation}
where $N$ is the number of states in the well being considered, $\hbar \wnm{n}{m}$ is the energy-level spacing between states \ket{n}\ and \ket{m}\ (tunable through $I_b$), and
\begin{equation}
  \Rnm{n}{m} = \frac{\Phio}{2 \pi} \frac{\Irf}{\hbar} \matel{n}{\gamma_1}{m}
\label{eqf}
\end{equation}
is a bare Rabi frequency.

The simplified Hamiltonian in \eqa{c}\ neglects tunneling through the potential barrier,\cite{Larkin86a, Leggett87a, Chow88a, Kopietz88a} a process corresponding to the SQUID spontaneously switching from the supercurrent to the voltage state; tunneling from \ket{n}\ occurs at an escape rate \Gn{n}.  As $\Gn{n+1} / \Gn{n} \sim 10^3$ for values of $I_b$ that we studied, tunneling can be exploited to perform state readout.  Note also typically $\Gn{n} \ll \wnm{n,}{n+1}$ for the lowest few levels; for states near the barrier, this is not true.  The total escape rate of the system to the voltage state is
\begin{equation}
  \Gamma = \frac{1}{\rho_{tot}} \sum_{n=0}^{N-1} \rho_{nn} \Gn{n},
\label{eqg}
\end{equation}
where $\rho_{nn}$ is the occupation probability of state \ket{n} and $\rho_{tot} = \sum \rho_{nn}$ is the probability of the system being in the supercurrent state; all of the quantities in the equation are time dependent during a bias ramp.

This description of the qubit makes several simplifications.  For one, we have ignored the quantum states of the isolation junction $J2$.\cite{Mitra06a}  For typical bias conditions, we keep the current through this junction small and its first excited state is two to three times higher in energy than that of the qubit.  However, the higher excited states of the qubit junction may come into resonance with states of the isolation junction.  This approach further ignores the role of $J2$ in determining the bias conditions.  Quantum mechanical simulations of the SQUID show that the current through the qubit junction weakly depends on its quantum state;\cite{Mitra06a} shifts in the qubit current are predicted to be less than 5 nA (much smaller than $I_{01} \approx 18\ \uA$) for our situation, so we have not considered this correction below.  As a result, the junction parameters in \eqa{a}\ need not be equal to the actual values for the qubit junction.

The simulations described in the sections that follow require values of \Gn{n}, \wnm{n}{m}, and \matel{n}{\gamma_1}{m}, all of which are specified by $I_b$, $I_{01}$, and $C_1$.  In the absence of dissipation, analytical expressions for these quantities can be obtained by applying perturbation theory to the cubic approximation of the tilted washboard potential.\cite{Strauch04a}  For the escape rates and energy levels, we instead use numerical solutions to the exact potential, which are more reliable for states near the top of the barrier.  These solutions come from two methods that give consistent results: (i) solving the Schr\"{o}dinger equation with transmission boundary conditions\cite{Shibata91a} and (ii) complex scaling.\cite{Moiseyev98a}

We performed experiments on a dc SQUID phase qubit fabricated by Hypres, Inc.\cite{NIST} using a $30\ \Acm$ Nb/AlO$_\text{x}$/Nb trilayer process [see Fig.~\ref{FDevice}(c)].  The qubit and isolation junctions had areas of $10 \times 10~\um^2$ and $7 \times 7~\um^2$, respectively, and the inductances were roughly $L_1 = 3.4\ \nH$, $L_2 = 30\ \pH$, and $M = 13.4\ \pH$.\cite{Palomaki07a}  The error in these and other fit parameters reported in this paper is about one unit in the least significant digit.  We have not attempted a thorough analysis of the uncertainties in all the parameters due to the complexities of the nonlinear functions involved.  The microwave current \Irf\ was carried inside the refrigerator on a single length of lossy stainless coax.  It was coupled to the qubit via an on-chip $1\ \fF$ capacitor \Crf\ [see Fig.~\ref{FDevice}(a)]; while the large impedance mismatch this produced potentially improved the isolation, it did make an independent calibration of the power reaching the qubit difficult.  The device was mounted in a superconducting aluminum box that was attached to the mixing chamber of a dilution refrigerator with a base temperature of 20 mK.  The refrigerator, located in an rf shielded room, was surrounded by a mu-metal cylinder.  In addition, the measurement and bias lines were protected from noise at room temperature by discrete $LC$ filters and copper powder filters at the mixing chamber.

The escape rate of the qubit junction was measured by simultaneously increasing $I_b$ and $I_f$ as described above, while monitoring the voltage on the current bias line of the SQUID.  We measured the time interval between the start of the ramps and when the system tunneled to the voltage state.  Repeating this procedure many times ($\sim 10^5$) at a rate of about 250 Hz yielded a histogram of switching times, from which the escape rate could be calculated.\cite{Fulton74a}  Decreasing the repetition rate did not yield a significant change in the escape rate, suggesting that heating due to the device switching to the voltage state had a minimal impact on the measurements.  Because the large loop inductance of the SQUID resulted in about 20 possible flux states, the device was initialized to the zero trapped flux state with a flux shaking procedure before each repetition of the measurement cycle.\cite{Palomaki06a}  We typically applied 50 oscillations of the flux current, which yielded a success rate better than 98\%.

\begin{figure}
  \vspace{0.1in}
  \includegraphics[bb=0 0 364 501, clip, width=3.0in]{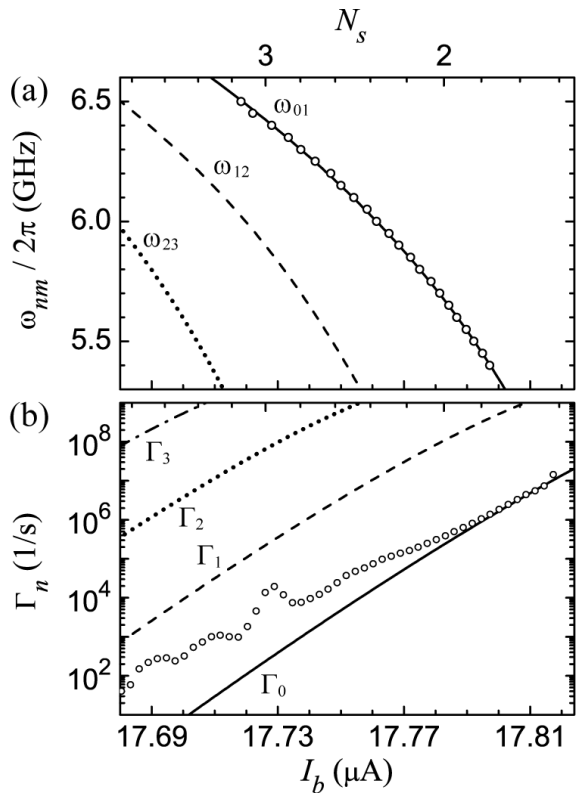}  \vspace{-0.1in}
  \caption{\label{FEnergyGamma} Qubit energy-level spectroscopy and tunneling escape rates.  (a) Open circles show the resonance frequency of the transition between the ground and first excited states of the qubit, measured at 20 mK.  The scatter in the values is indicative of the uncertainty in the measurement.  Also plotted are theoretical values of \wnm{0}{1}\ (solid), \wnm{1}{2}\ (dashed), and \wnm{2}{3}\ (dotted) for $I_{01} = 17.930\ \uA$ and $C_1 = 4.50\ \pF$.  (b) At high bias, the measured background escape rate (open circles) agrees with the predicted ground-state escape rate \Gn{0}\ (solid line) for the junction parameters given above.  Calculated \Gn{1}, \Gn{2}, and \Gn{3}\ are plotted as dashed, dotted, and dashed-dotted lines.  In both plots, the bottom axes show the total bias current $I_b$, while the top axes indicate the normalized barrier height $N_s$, calculated using the extracted junction parameters.}
\end{figure}

Figure~\ref{FEnergyGamma}(a) shows a spectrum of transitions for the qubit junction.  With a microwave drive of fixed frequency applied to the device, the total escape rate was measured while ramping the biases.  The open circles indicate the values of $I_b$ where the microwave drive caused the largest enhancement in the escape rate (over its background value in the absence of microwaves).

As shown by the solid line, we fit these points to the theoretical values of \wnm{0}{1}, yielding $I_{01} = 17.930\ \uA$ and $C_1 = 4.50\ \pF$.  These values are likely to differ from the actual critical current and capacitance of the qubit junction, both in principle, due to the simplifications in the Hamiltonian mentioned above, and in practice, as this fitting procedure is sensitive to inaccuracies in the simultaneous ramping of $I_b$ and $I_f$, constant offset flux that biases the SQUID (from, for example, trapped flux near the device), and drifts in the detection electronics.  Nonetheless, when we repeated the measurement at elevated temperature, the location of the transitions between higher levels agreed to within 0.2\% of predictions for \wnm{1}{2}\ (dashed) and \wnm{2}{3}\ (dotted) obtained with the effective parameters.\cite{Strauch06a}  In addition, two-photon transitions between these levels occurred at high power at the expected frequencies.  Multiphoton transitions between the ground and first excited states of a variety of junction qubits have previously been observed and characterized.\cite{Nakamura01a, Wallraff03a, Saito04a}  Over the course of three months (during which the refrigerator remained near its base temperature), the junction parameters found from the spectral fits varied by roughly 1\%.  Thus, we repeated the spectroscopy measurement often to obtain the values of $I_{01}$ and $C_1$ needed for the simulations; these values are listed in the caption of each figure.

The open circles of Fig.~\ref{FEnergyGamma}(b) show the measured background escape rate in the absence of microwaves.  The solid line indicates the theoretical value of \Gn{0}, calculated with the values of $I_{01}$ and $C_1$ extracted from the spectrum.  At low bias current, the measured escape rate exceeds \Gn{0}, suggesting the presence of excited state population even though the refrigerator was at 20 mK.  Based on experiments described elsewhere,\cite{Palomaki07a} we believe that features such as the peak near $I_b = 17.73\ \uA$ are due to population in \ket{2}\ generated when noise on the leads at the resonant frequency of the isolation junction matches the \trans{0}{2}\ transition frequency of the qubit.  At high bias, the escape rates exceed the excitation rates of the noise and the total escape rate converges to $\Gamma_0$; a similar effect occurs for thermal excitations.\cite{Dutta04a}  While noise complicates the situation, the overall behavior suggests that both \wnm{0}{1}\ and \Gn{0}\ of the dc SQUID are described by the same one-dimensional tilted washboard potential.  Also shown in Fig.~\ref{FEnergyGamma}(b) are numerical predictions for \Gn{1}\ (dashed), \Gn{2}\ (dotted), and \Gn{3}\ (dashed-dotted).

\begin{figure}
  \includegraphics[bb=0 0 383 473, clip, width=3.0in]{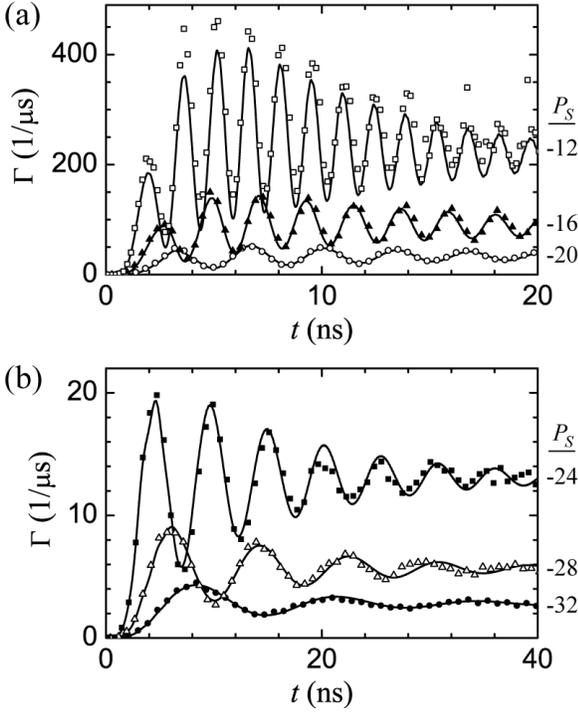}    \vspace{-0.1in}
  \caption{\label{F031806DGtot} Rabi oscillations in the escape rate $\Gamma$ were induced at $I_b = 17.746\ \uA$ by switching on a microwave current at $t = 0$ with a frequency of 6.2 GHz (resonant with the \trans{0}{1}\ transition) and source powers \Ps\ between $-12$ and $-32$ dBm, as labeled.  The measurements were taken at 20 mK.  The solid lines are from a five-level density-matrix simulation with $I_{01} = 17.930\ \uA$, $C_1 = 4.50\ \pF$, $T_1 = 17\ \ns$, and $T_\phi = 16\ \ns$.}
\end{figure}

We observed Rabi oscillations between the ground and first excited states by turning the microwave current \Irf\ on when $I_b$ was at the value where $\wnm{0}{1} = \wrf$ and measuring the time-resolved escape rate while \Irf\ remained at a constant value.  This serves as a simple method of monitoring the evolution of the state populations.\cite{Yu02a}  Although the biases continued to increase during the Rabi oscillations, the ramp rates were reduced (with $d I_b / dt \approx 0.01\ \As$) before \Irf\ was turned on, so that the level spacing $\wnm{0}{1}$ changed by a negligible amount during the escape rate measurement.  The symbols in Fig.~\ref{F031806DGtot} show $\Gamma$ due to a 6.2 GHz microwave drive for a range of powers $\Ps$ at the microwave generator.  As expected, the oscillation frequency increases with power and decoherence causes the amplitude of the oscillations to decay with time.

However, there are two unexpected features in the data.  First, the escape rate \Geq\ at long time (once the oscillations have decayed away) increases with \Ps, whereas for an ideal two-level system, the excited state population saturates at high power.  Furthermore, the highest measured escape rates far exceed the value of $2.2 \times 10^6\ \invs$ predicted for \Gn{1}\ at this bias current, strongly suggesting that the states \ket{2}\ and higher are becoming occupied.  As the escape rates from these levels are very large, only a small population would be required to account for the observed $\Gamma$.  Second, at high power, the oscillation maxima increase over the first few cycles [see Fig.~\ref{F031806DGtot}(a)]; this is due to the rise time of the microwave current amplitude, which will be discussed in Sec.~\ref{SDecoherence}.

In Sec.~\ref{SStrong}, we will show how more complete measurements reveal that the higher excited states impact the Rabi oscillation frequencies in ways that are in quantitative agreement with predictions from the Hamiltonian of \eqa{c}, but are difficult to see in Fig.~\ref{F031806DGtot}.  As shown by the solid lines in the figure, most of the features of the measured escape rate are captured by a multilevel density-matrix simulation which will be described in Sec.~\ref{SDecoherence}.

\section{Detuning and Strong Field Effects}
\label{SStrong}

We first model Rabi oscillations using the rotating wave approximation, in the absence of tunneling and dissipation.  This provides a way of predicting the oscillation frequency for a wide range of experimental parameters.\cite{Steffen03a, Claudon04a, Meier05a, Amin06a, Strauch06a, Claudon07b}  In the rotating frame corresponding to the drive frequency $\wrf \approx \wnm{0}{1}$, the Hamiltonian of \eqa{c}\ simplifies to\cite{Strauch06a}
\begin{equation}
  \Ham_N^{RW} = \sum_{n=0}^{N-1} \hbar \Delta_n \ket{n} \bra{n}
             + \frac{1}{2} \sum_{\substack{n,m=0 \\ n \neq m}}^{N-1}
               \hbar \Omega_{nm}^\prime \ket{n} \bra{m},
\label{eqd}
\end{equation}
where $\Delta_n = \wnm{0}{n} - n \wrf$ and
\begin{equation}
  \Omega_{nm}^\prime = \Rnm{n}{m} \sum_{s=0}^1 J_{m-n+2s-1}
                       \paren{\frac{\Rnm{n}{n} - \Rnm{m}{m}}{\wrf}}
\end{equation}
for $n < m$ and $\Omega_{mn}^\prime = \Omega_{nm}^\prime$.  Here, $J_n \paren{x}$ is the $n$th order Bessel function and $\Omega_{nm}$ is defined by \eqa{f}.

Differences between the $N$ eigenvalues of this Hamiltonian specify effective multilevel Rabi frequencies or modes of the system.  We find it convenient to label these frequency differences by the states \ket{n}\ with the largest weight in the two corresponding eigenfunctions.  For the parameter regime of interest here, the differences can be uniquely classified by two states, which we denote by \RReffnm{n}{m}.  For example, \RReffnm{0}{2}\ denotes the Rabi oscillation frequency between eigenstates that describe a two-photon transition between \ket{0}\ and \ket{2}.  While approximate analytical solutions can be obtained for a three-level junction system,\cite{Strauch06a} we numerically found the eigenvalues of the rotating wave Hamiltonian for systems with up to seven levels in the simulations that follow.

\begin{figure}
  \includegraphics[bb=0 0 369 263, clip, width=3.0in]{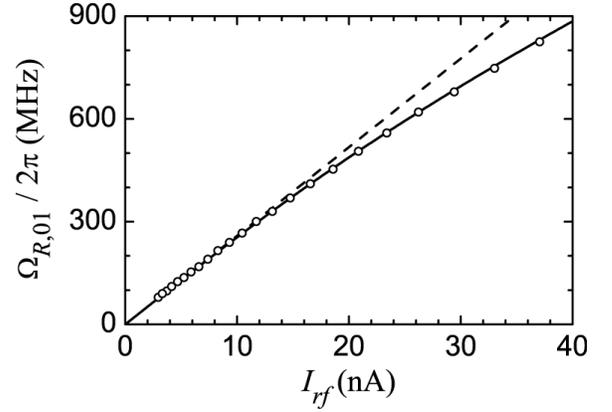}    \vspace{-0.1in}
  \caption{\label{F031806Dfosc} Rabi oscillation frequency \RReffnm{0}{1}\ at fixed bias as a function of microwave current \Irf.  Extracted values from data (including the plots in Fig.~\ref{F031806DGtot}) are shown as circles, while the rotating wave solution is shown for two- (dashed line) and five- (solid) level simulations, calculated using $I_{01} = 17.930\ \uA$ and $C_1 = 4.50\ \pF$ with $\wrf / 2 \pi = 6.2\ \GHz$.}
\end{figure}

We fit the escape rates in Fig.~\ref{F031806DGtot} (and additional data for other powers not shown) to a decaying sinusoid with an offset.  The extracted frequencies are shown with circles in Fig.~\ref{F031806Dfosc}.  To compare to theory, \RReffnm{0}{1}, calculated using the rotating wave solution for a system with five levels, is shown with a solid line.  The implied assumption that the oscillation frequencies of $\Gamma$ and $\rho_{11}$ are equal, even at high power in a multilevel system, will be addressed in Sec.~\ref{SSummary}.  In plotting the data, we have introduced a single fitting parameter $117\ \AW$ that converts the power \Ps\ at the microwave source to the current amplitude \Irf\ at the qubit.  Good agreement is found over the full range of power.

As \Irf\ increases in Fig.~\ref{F031806Dfosc}, the oscillation frequency is smaller than the expected linear relationship for a two-level system (dashed line).  This effect is a hallmark of a multilevel system and has been previously observed in a similar phase qubit.\cite{Claudon04a, Claudon07b}  There are two distinct phenomena that affect \trans{0}{1}\ Rabi oscillations in such a device.\cite{Goorden03a, Meier05a, Strauch06a, Shevchenko07a}  To describe these, we must first define resonance as occurring when the Rabi frequency \RReffnm{0}{1}\ is at a minimum, as the detuning between the microwave drive frequency \wrf\ and level spacing \wnm{0}{1}\ is varied (for a fixed drive power).   In a two-level system, this happens for $\wnm{0}{1} = \wrf$.  For our phase qubit, resonance occurs when $\wnm{0}{1} < \wrf$ due to the decreasing level spacings with increasing state \ket{n}\ [see Fig.~\ref{FEnergyGamma}(a)].  This shift in resonance is an analog of the ac Stark effect.  Resonance shifts occur under strong driving in other superconducting qubits as well.\cite{Schuster05a, Dutton06a}

However, this shift does not explain the data shown in Fig.~\ref{F031806Dfosc}, which were taken at a fixed bias current (and thus off resonance at high power).  The higher levels also affect the frequency of the Rabi oscillations. We will refer to the minimum value of \RReffnm{n}{m}\ as the on-resonance Rabi frequency \RRnm{n}{m}.  The suppression of \RRnm{0}{1}\ below \Rnm{0}{1}\ leads to the effect seen in Fig.~\ref{F031806Dfosc}.

Both of these effects become significant as \Rnm{0}{1}\ approaches $\wnm{0}{1} - \wnm{1}{2}$, which is a measure of the anharmonicity of the system;\cite{Meier05a, Amin06a, Strauch06a} in the case of Fig.~\ref{F031806Dfosc}, $\wnm{0}{1} / 2 \pi = 6.2\ \GHz$ and $\wnm{1}{2} / 2 \pi = 5.5\ \GHz$.  Clearly, these shifts need to be considered when working at high power or at low current bias.

\begin{figure}
  \includegraphics[bb=0 0 461 539, clip, width=3.0in]{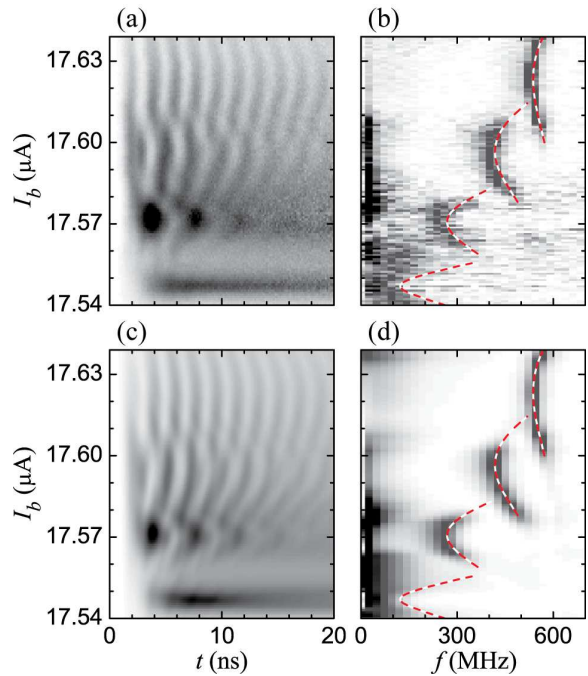}    \vspace{-0.1in}
  \caption{\label{F010206H1} (Color online) Multiphoton, multilevel Rabi oscillations plotted in the time and frequency domains.  (a) The escape rate $\Gamma$ (measured at 110 mK) is plotted as a function of the time after which a 6.5 GHz, $-11$ dBm microwave drive was turned on and the current bias $I_b$ of the qubit; $\Gamma$ ranges from 0 (white) to $3 \times 10^8\ \invs$ (black).  (b) The normalized power spectral density of the time-domain data from $t = 1$ to 45 ns is shown with a grayscale plot.  The dashed line segments indicate the Rabi frequencies obtained from the rotating wave model for transitions involving (from top to bottom) 1, 2, 3, and 4 photons, evaluated with junction parameters $I_{01} = 17.828\ \uA$ and $C_1 = 4.52\ \pF$, and microwave current $\Irf = 24.4\ \nA$.  Corresponding grayscale plots calculated with a seven-level density-matrix simulation are shown in (c) and (d).}
\end{figure}

In order to follow experimentally the shift of the resonance condition, it was necessary to measure Rabi oscillations for different detunings of the microwave drive.  We chose to do this by keeping the drive frequency \wrf\ fixed and changing the level spacing \wnm{0}{1}\ (through $I_b$), because the power transmitted by the microwave lines had a nontrivial frequency dependence.  Figure~\ref{F010206H1}(a) shows a grayscale plot of Rabi oscillations measured from such an experiment, where black represents a high escape rate.  Each horizontal line is the escape rate versus time due to a microwave current of 6.5 GHz and $-11$ dBm, which was turned on at the value of the current bias $I_b$ indicated on the vertical axis.  While the measurements were performed at 110 mK, this is not expected to have a significant impact on the Rabi oscillations, as the temperature was well below $\hbar \wnm{0}{1} / k_B \approx 325\ \mK$.\cite{Lisenfeld07a}

From Fig.~\ref{F010206H1}(a), we see that the oscillation frequency depends on the current bias.  This variation can be seen more readily in Fig.~\ref{F010206H1}(b), which shows the power spectral density (calculated as the absolute square of the discrete Fourier transform) of the escape rate data in Fig.~\ref{F010206H1}(a)  between $t = 1$ and $45\ \ns$.  For this plot, each horizontal line has been normalized to its maximum value (black) in order to emphasize the location of the dominant frequency.  Three distinct bands are visible.

For this data set, the level spacing $\wnm{0}{1} / 2 \pi$ is equal to the microwave frequency $\wrf / 2 \pi = 6.5\ \GHz$ at $I_b = 17.614\ \uA$.  The band with the highest current in Fig.~\ref{F010206H1}(b) is centered about $I_b = 17.624\ \uA$, suggesting that \trans{0}{1}\ Rabi oscillations are the dominant process near this bias.  For slightly higher or lower $I_b$, the oscillation frequency increases as $\RReffnm{0}{1} \approx \sqrt{\Omega_{01}^{\prime2} + \paren{\wrf - \wnm{0}{1}}^2}$, in agreement with simple two-level Rabi theory, leading to the curved band in the grayscale plot.

The other bands correspond to oscillations between the ground state and higher excited states.  Due to the anharmonic level structure, $\wnm{0}{2} / 4 \pi$ is 6.5 GHz at a smaller current bias $I_b = 17.594\ \uA$; thus a second band appears there, corresponding to two-photon \trans{0}{2}\ Rabi oscillations.  Similarly, three-photon \trans{0}{3}\ oscillations are visible near $17.572\ \uA$ where $\wnm{0}{3} / 6 \pi = 6.5\ \GHz$.  Finally, large escape rates occur near $17.549\ \uA$ corresponding to a four-photon \trans{0}{4}\ transition, although no oscillations are apparent in Fig.~\ref{F010206H1}(a).  Note that the effective junction parameters $I_{01}$ and $C_1$ (given in the caption) used to predict \wnm{n}{m}\ and other level properties are slightly different than those for Figs.~\ref{FEnergyGamma}--\ref{F031806Dfosc}, as the data sets were taken two months apart.

The rotating wave solution provides a simple way to predict the oscillation frequencies.  Calculations of \RReffnm{0}{n}\ using \eqa{d}\ for a seven-level system are shown as dashed curves in Fig.~\ref{F010206H1}(b) for the four lowest multiphoton transitions ($n = 1, 2, 3, 4$).  The microwave amplitude $\Irf = 24.4\ \nA$ is the only free parameter in the calculation and the rotating wave solution using this value reproduces the oscillation frequencies of the different processes well, even at large detuning.

Figure~\ref{F010206H1}(b) shows that the minimum oscillation frequency $\RRnm{0}{1} / 2 \pi = 540\ \MHz$ of the first (experimental) band occurs at $I_b = 17.624\ \uA$, for which $\wnm{0}{1} / 2 \pi = 6.4\ \GHz$.  This again indicates an ac Stark shift of this transition, which we denote by $\Delta \wnm{0}{1} \equiv \wrf - \wnm{0}{1} \approx 2 \pi \times 100\ \MHz$.  In addition, the higher levels have suppressed the oscillation frequency below the bare Rabi frequency of $\Rnm{0}{1} / 2 \pi = 620\ \MHz$ [calculated with \eqa{f}].

\begin{figure}
  \includegraphics[bb=0 0 375 461, clip, width=3.0in]{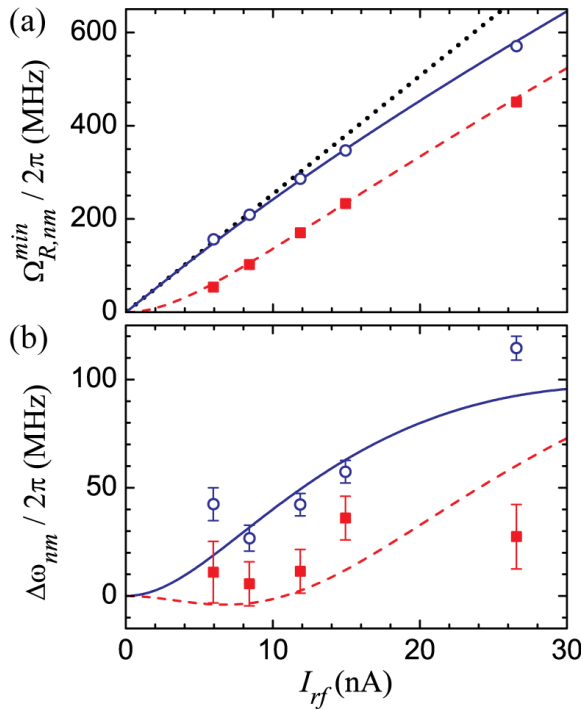}    \vspace{-0.1in}
  \caption{\label{F032206MNstats} (Color online) The (a) on-resonance Rabi oscillation frequencies \RRnm{0}{1}\ and \RRnm{0}{2}\ and (b) resonance frequency shifts $\Delta \wnm{0}{1} = \wrf - \wnm{0}{1}$ and $\Delta \wnm{0}{2} = 2 \wrf - \wnm{0}{2}$ are plotted as a function of the microwave current, for data taken at $110\ \mK$ with a microwave drive of frequency $\wrf / 2 \pi = 6.5\ \GHz$ and powers $\Ps = -23, -20, -17, -15, -10\ \dBm$.  Values extracted from data for the \trans{0}{1}\ (\trans{0}{2}) transition are plotted as open circles (filled squares), while five-level rotating wave solutions for a junction with $I_{01} = 17.736\ \uA$ and $C_1 = 4.49\ \pF$ are shown as solid (dashed) lines.  In (a), the dotted line is from a simulation of a two-level system.}
\end{figure}

We repeated this analysis of resonance for five different microwave powers, with data taken at a later date.  Figure~\ref{F032206MNstats} shows experimental results for the on-resonance Rabi frequencies $\RRnm{n}{m}$ and Stark shifts $\Delta \wnm{n}{m}$ with the corresponding results from a five-level rotating wave solution for the \trans{0}{1}\ (circles for data and solid lines for theory) and two-photon \trans{0}{2}\ (squares and dashed lines) transitions.  Here, the power calibration is $84\ \AW$ for $\wrf / 2 \pi = 6.5\ \GHz$.  Figure~\ref{F032206MNstats}(a) differs from Fig.~\ref{F031806Dfosc}, because in the former, $I_b$ was varied at each power to give the minimum oscillation frequency; by staying on resonance in this way, the effect of the higher levels on the \trans{0}{1}\ oscillations is maximized.

The resonant oscillation frequencies \RRnm{0}{1}\ and \RRnm{0}{2}\ in Fig.~\ref{F032206MNstats}(a) are well described by \eqa{d} over the full range of \Irf.  The deviation between the \trans{0}{1}\ oscillation frequency and the values expected in a two-level system (dotted line) increases with \Irf. Similar measurements taken at a higher $I_b$ show a smaller frequency suppression over a similar range of \Irf,\cite{Strauch06a} as expected for a system with stronger anharmonicity.

The resonance shifts $\Delta \wnm{0}{1}$ and $\Delta \wnm{0}{2}$ in Fig.~\ref{F032206MNstats}(b), however, differ significantly from the model predictions.  As a change of 10 MHz in $\wnm{0}{1} / 2 \pi$ corresponds to a roughly 1 nA change in $I_b$, the discrepancy is difficult to see in Fig.~\ref{F010206H1}(b).  The indicated uncertainty in the experimental points in Fig.~\ref{F032206MNstats}(b) is roughly 5 MHz due to errors in the calibration of $\wnm{0}{1} \paren{I_b}$.  In addition the uncertainty is somewhat larger at low power, where the relatively small total escape rates result in poor counting statistics, and high power, where the weak dependence on detuning makes it difficult to identify the resonant level spacing.  While the general trend of the shifts is consistent with the model, the scatter in the data is large and further work would be needed to determine if there are true deviations from the multilevel theory.

\section{Decoherence}
\label{SDecoherence}

We now model the time dependence of the escape rate measured in the experiment.  In order to do this, several additions have to be made to the treatment of the system given in Sec.~\ref{SStrong}: the effects of tunneling, other sources of damping and noise, and experimental limitations.  The density-matrix formalism\cite{Smith78a, Blum96a} provides a straightforward scheme for including nonunitary processes and has been previously applied to the lowest two or three levels of the phase qubit.\cite{Yu02a, Kosugi05a, Meier05a, Amin06a, Claudon07b, Shevchenko07a}  For a system with $N$ levels, we assume that the evolution of the qubit's reduced density matrix \dm\ is given by the modified Liouville--von Neumann equation
\begin{equation}
  \partiald{\dm}{t} = -\frac{i}{\hbar} \brac{\Ham_N, \dm} - G \dm
                          - R \dm - D \dm.
\label{eqb}
\end{equation}
Here, we use the discrete Hamiltonian in \eqa{c}, so that the diagonal elements $\rho_{nn}$ give the occupancy of the states \ket{n}.  Tunneling is characterized by the tensor $G$, where $\brac{G \dm}_{nm} = \paren{\Gn{n} + \Gn{m}} \rho_{nm} / 2$, which leads to a decay of all elements of \dm.  The tensors $R$ and $D$ account for two distinct decoherence mechanisms.

The Bloch-Redfield tensor $R$ models the effects of the system being in equilibrium with a thermal bath at temperature $T$.\cite{Burkard04a}  This tensor leads to the decay of the diagonal elements of $\rho$ (dissipation) as well as the off-diagonal elements (decoherence).  The coupling to the bath, assumed linear in $\gamma_1$, is parametrized by $R_1 \paren{\omega}$, which is the inverse of the real part of the total admittance that shunts the qubit junction, evaluated at angular frequency $\omega$.  In \eqa{b},
\begin{equation}
  \brac{R \dm}_{nm} = \sum_{k,l=0}^{N-1} R_{nmkl} \rho_{kl},
\label{eqe}
\end{equation}
where
\begin{equation}
  R_{nmkl} = - \gamma_{lmnk} - \gamma_{knml}
             + \delta_{lm} \sum_{r=0}^{N-1} \gamma_{nrrk}
             + \delta_{nk} \sum_{r=0}^{N-1} \gamma_{mrrl}
\end{equation}
and
\begin{widetext}
\begin{equation}
  \gamma_{lmnk} = \frac{1}{2 \hbar} \paren{\frac{\Phio}{2 \pi}}^2
      \frac{\matel{l}{\gamma_1}{m} \matel{n}{\gamma_1}{k}}
      {R_1 \paren{\wnm{n}{k}}}
      \brac{\paren{1 - \delta_{nk}} \wnm{n}{k}
      \frac{\exp \paren{-\frac{\hbar \wnm{n}{k} \mathrm{sgn} \paren{n - k}}
             {2 k_B T}}}
           {\sinh \paren{\hbar \wnm{n}{k} / 2 k_B T}}
      + \delta_{nk} \frac{2 k_B T}{\hbar}}.
\end{equation}
\end{widetext}
In the following, we assume that $R_1$ is independent of frequency, in which case we can define a dissipation time $T_1 = R_1 C_1$.  In this limit, \eqa{e} gives well known interlevel transition rates.\cite{Larkin86a, Chow88a, Xu05b}  For example, thermal excitation from \ket{n}\ to \ket{m}\ (with $m > n$) occurs at a rate
\begin{eqnarray}
  W_{mn}^+ & = & 2 \gamma_{nmmn}  \nonumber \\
           & = & \frac{2}{\hbar} \paren{\frac{\Phio}{2 \pi}}^2
                 \paren{\frac{\wnm{n}{m}}{R_1}} \frac{\abs{\matel{n}{\gamma_1}{m}}^2}
                 {\exp \paren{\hbar \wnm{n}{m} / k_B T} - 1}.
\end{eqnarray}
As required by detailed balance, decay from \ket{m}\ to \ket{n}\ occurs at a rate $W_{nm}^- = W_{mn}^+ \exp \paren{\hbar \wnm{n}{m} / k_B T}$.  In addition, \eqa{e} specifies the decoherence due to this dissipation.  Given $\wnm{n}{m}$ and the matrix elements of $\gamma_1$, all of the thermal rates are specified by $T_1$, which is set to $17\ \ns$ for the simulations below.  This value comes from additional measurements of the escape rate of the device in the absence of microwaves over a range of temperatures.\cite{Dutta04a}

We find that this treatment of dissipation alone is insufficient to capture the decay of Rabi oscillations, suggesting that an additional decoherence mechanism is present.  This we model by the tensor $D$ in \eqa{b}, which has the form
\begin{equation}
  D \dm = \sum_n \lambda_n \paren{L_n \dm L_n^\dag
        - \frac{1}{2} L_n^\dag L_n \dm - \frac{1}{2} \dm L_n^\dag L_n},
\label{eqh}
\end{equation}
where $L_n$ are Lindblad operators with strengths $\lambda_n$.  The best overall agreement with the measurements (see further discussion below) is found for a set of operators $L_n = \ket{n} \bra{n}$, where $\lambda_n = 1 / T_\phi$ and $n$ ranges from 0 to $N-1$.  This leads to an exponential decay of each of the off-diagonal elements of \dm\ with a common time constant given by the dephasing time $T_\phi$ and no change in the diagonal elements.  In the simulations of this section, we set $T_\phi = 16\ \ns$, yielding a coherence time $T_2 = \brac{1 / \paren{2 T_1} + 1 / T_\phi}^{-1} = 10.9\ \ns$.

For $N = 2$, \eqa{b}\ reduces to the optical Bloch equations,\cite{Smith78a, Blum96a}  for which analytical solutions exist.  However, for $N > 2$ the master equation is easily numerically integrated to obtain the time dependence of \dm, without making the rotating wave approximation.  With the state occupation probabilities $\rho_{nn}$ in hand, the total escape rate can be calculated with \eqa{g}\ and compared with experiment.

\begin{figure}
  \includegraphics[bb=0 0 373 468, clip, width=3.0in]{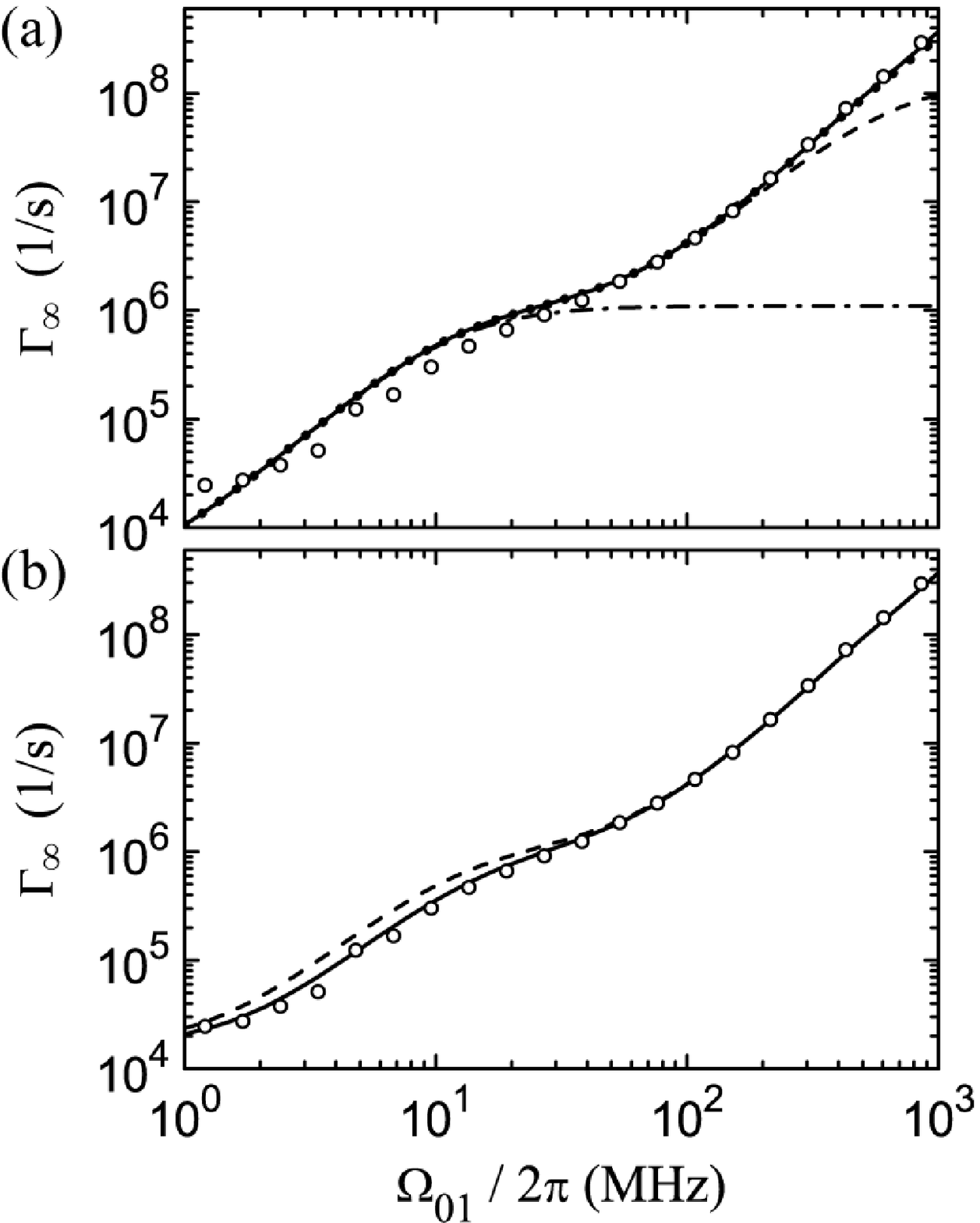}    \vspace{-0.1in}
  \caption{\label{F031806DGeq} Long-time escape rate \Geq\ as a function of the bare Rabi frequency \Rnm{0}{1}.  (a) \Geq\ is plotted for a density-matrix simulation of a system with two (dashed-dotted), three (dashed), four (dotted), and five (solid) levels, $I_{01} = 17.930\ \uA$, and $C_1 = 4.50\ \pF$.  (b) \Geq\ at low power (for a five-level system) is affected by the inclusion of a small microwave current at \wnm{0}{2}\ (dashed) and inhomogeneous broadening in addition to this noise current (solid).  One set of experimental data, for source powers between $-68$ and $-11\ \dBm$ at 6.2 GHz, is plotted in both panels with open circles.}
\end{figure}

As mentioned earlier, one of the more striking features of the series of \trans{0}{1}\ Rabi oscillations in Fig.~\ref{F031806DGtot} is that \Geq, which we define as the steady escape rate the system approaches as the oscillations decay away, increases over the full range of measured power.  The circles in Fig.~\ref{F031806DGeq} show experimental values of \Geq\ as a function of the bare Rabi frequency \Rnm{0}{1}, which was calculated from the microwave source power by using the fit in Fig.~\ref{F031806Dfosc}.  We will now use the density-matrix simulations to understand these escape rates.

The simulations suggest that \Geq\ is not sensitive to the time evolution of \dm, and is only weakly dependent on $T_1$ and $T_\phi$ at high power.  However, \Geq\ does depend on the number of states that participate in the dynamics and their individual escape rates \Gn{n}.  It should be noted that the actual populations do not reach steady state, but continuously decay due to tunneling.  The dashed-dotted line in Fig.~\ref{F031806DGeq}(a) shows the calculated \Geq\ for a two-level system ($N = 2$).  It increases up to a value of $\Gn{1} / 2$ at roughly $\Rnm{0}{1} = 1 / \sqrt{T_1 T_2}$; the data show only a subtle shoulder, masked by an overall steady increase, near this escape rate.  Results for three (dashed), four (dotted), and five (solid) levels are also plotted in the figure, displaying increasing agreement with experiment.

This agreement between theory and experiment at high power is an indication that transitions to the higher states, or leakage in the context of quantum computation, are occurring as expected.  As power increases, the occupation of higher excited states increases and this effect is magnified by their larger escape rates.  For example, with the five-level simulation at $\Rnm{0}{1} / 2 \pi = 1\ \GHz$ the state \ket{3}\ has a 2\% occupation probability, but accounts for 60\% of \Geq.  Note that microwave pulse shaping will not reduce this effect.  That is, while leakage can be minimized at the end of a single qubit operation, higher states are always populated during the pulse.\cite{Steffen03a, Meier05a, Amin06a}  Such pulses will have enhanced escape rates comparable to those seen here for a constant \Irf.

These simulations do not do a good job of explaining the value of \Geq\ at very low power.  The reason for this can be seen in Fig.~\ref{FEnergyGamma}(b), where the measured escape rate in the absence of microwaves exceeds the predicted value of \Gn{0}.  In particular, at the bias current of $I_b = 17.746\ \uA$ where the Rabi oscillations were performed, the measured escape rate is $\sim 2 \times 10^4\ \invs$.  While this excess escape rate could be attributed to a thermal population at $T = 56\ \mK$ (whereas the refrigerator thermometer read 20 mK), raising the temperature of the simulation leads to a significant enhancement in \Geq\ that extends up to moderate microwave power, an effect not seen in the data.  Instead, separate measurements\cite{Palomaki07a} indicate that the spurious features in the measured $\Gamma$ are largely due to a population in \ket{2}, beyond that expected from a thermal bath at 20 mK.  Assuming that this was a result of noise on the bias lines, we included in the simulation an additional microwave source at a frequency of \wnm{0}{2}; a microwave current amplitude of 0.2 nA reproduces the background escape rate measured at $I_b = 17.746\ \uA$ (corresponding to Figs.\ \ref{F031806DGtot}, \ref{F031806Dfosc}, and \ref{F031806DGeq}) and leads to a population $\rho_{22} = 3 \times 10^{-5}$.  The coherent effects of this drive are negligible, as $\Rnm{0}{2} / 2 \pi = 0.5\ \MHz$, which is much smaller than $1/T_1$ or $1/T_\phi$.  The results of a five-level simulation with the extra source included are shown with a dashed line in Fig.~\ref{F031806DGeq}(b).  Compared to the simpler simulations in Fig.~\ref{F031806DGeq}(a), they show improved agreement at the lowest powers, with little change above $\Rnm{0}{1} / 2 \pi = 10\ \MHz$.

To this point, we have not included effects from inhomogeneous broadening, although our spectroscopic measurements suggest its presence.  For $T_2 = 10.9\ \ns$, which describes many of the experiments below, the expected full width at half maximum of a resonance peak is roughly 30 MHz.  However, we commonly find peak widths of 50 MHz at bias currents where tunneling makes a negligible contribution.  The remaining broadening may be due to current noise at frequencies much lower than $1 / T_1$, whose contribution to the width scales with the spectral slope $d \wnm{0}{1} / d I_b$.\cite{Berkley03b}  This can be modeled by taking into account the frequency content of the noise and using the stochastic Bloch equations.\cite{Xu05b}  Instead, we mimicked the inferred spectroscopic broadening simply by running the simulation for a range of bias currents and then convolving the resulting escape rate (at a given time) and a Gaussian with standard deviation $\sigma_I = 1.5\ \nA$ (corresponding to a 35 MHz spread in $\wnm{0}{1} / 2 \pi$).

Calculations of \Geq\ for a five-level system with noise at \wnm{0}{2}\ and this inhomogeneous broadening are drawn with a solid line in Fig.~\ref{F031806DGeq}(b).  The extra broadening has negligible effect at high power (where Rabi oscillations are observed), but does bring the simulation into better agreement with data near $\Rnm{0}{1} / 2 \pi = 10\ \MHz$.  The small remaining discrepancy could be due to detuning arising from a misidentification of the mean value of $I_b$.  For example, performing the simulation at 17.745 rather than $17.746\ \uA$ results in an underestimate of the measured values.  Nevertheless, the overall agreement over three decades in the Rabi frequency is very good.  While the calibration of the microwave current \Irf\ came from a fit to data, one conversion factor reproduces both the oscillation frequencies (Fig.~\ref{F031806Dfosc}) and long-time escape rates (Fig.~\ref{F031806DGeq}).

\begin{figure}
  \includegraphics[bb=0 96 360 649, clip, width=3.0in]{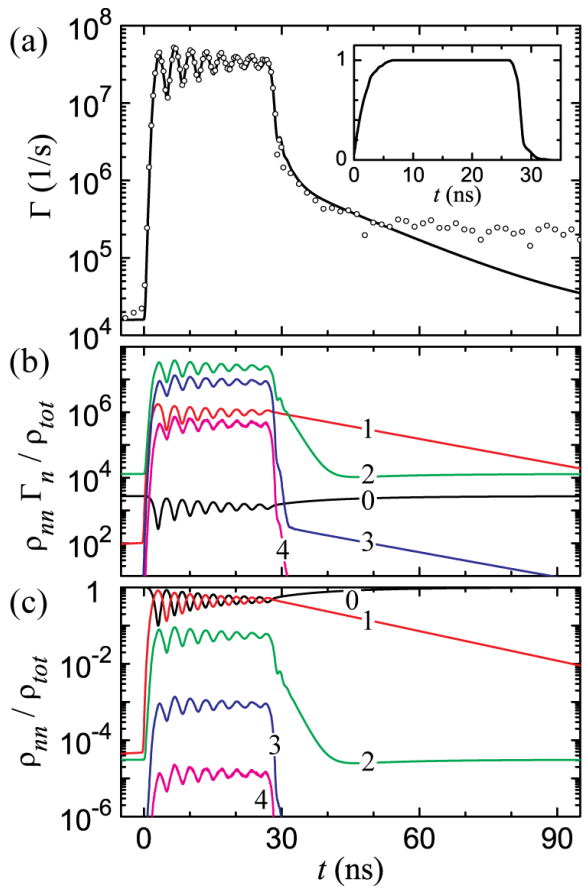}    \vspace{-0.0in}
  \caption{\label{F032106FN10} (Color online) Response to a microwave pulse.  (a) The measured escape rate $\Gamma$ (circles), due to a 6.2 GHz, $-20\ \dBm$ microwave pulse nominally 30 ns long, shows Rabi oscillations followed by a decay governed by multiple time constants.  The solid line is the result of a five-level density-matrix simulation for a junction with $I_{01} = 17.730\ \uA$ and $C_1 = 4.46\ \pF$, using the measured microwave pulse amplitude (with a maximum of 11.7 nA) shown in the inset.  The simulation also produces (b) the contribution to the escape rate and (c) the normalized occupation for each of the five levels, as labeled.}
\end{figure}

We next consider the time dependence of the escape rate for the data plotted in Fig.~\ref{F032106FN10}.  Here, a 6.2 GHz microwave pulse nominally 30 ns long was applied on resonance with the \trans{0}{1}\ transition of the qubit junction.  The measured escape rate shows Rabi oscillations followed by a decay back to the ground state once the microwave drive has turned off.  This decay appears to be governed by three time constants.  Nontrivial decays have previously been reported in phase qubits\cite{Martinis02a} and we have found them in several of our devices.

Accurately simulating this experiment requires knowledge of the time dependence of the microwave pulse, which was created by the internal gate of a Hewlett-Packard 83731B synthesized source,\cite{NIST} without any further filtering.  We measured the pulse envelope at the source's output using a digital sampling oscilloscope [see inset of Fig.~\ref{F032106FN10}(a)].  Ignoring any distortion of the pulse before it reached the qubit junction, this was taken as the microwave amplitude $\Irf \paren{t}$.  For the microwave power used in Fig.~\ref{F032106FN10}, the maximum value of \Irf\ was taken to be 11.7 nA (a value obtained from the power calibration in Fig.~\ref{F031806Dfosc}). The solid line in Fig.~\ref{F032106FN10}(a) shows the calculated escape rate for a five-level simulation in the presence of noise at \wnm{0}{2}\ and inhomogeneous broadening.  For this time-domain plot and others discussed below, we have convolved the simulation and a Gaussian with a full width of 150 ps to remove very fast, small oscillations (due to highly detuned multiphoton processes) that could not be seen in the experiment due to insufficient time resolution.

The main part of the Rabi oscillation is reproduced well.  In particular, the second oscillation maximum has a larger escape rate than the first, due to the 7 ns it takes for \Irf\ to reach its maximum.  The first part of the decay is also reproduced, which from the simulation should correspond to the emptying of state \ket{2}.  However, the data also show a slow decay constant longer than 50 ns that the simulation does not explain.  The longer time is inconsistent with our thermal measurement of $T_1$ and instead may be indicative of the qubit interacting with an additional quantum system.  It appears, though, that this extra degree of freedom does not significantly affect the description of the Rabi oscillations.  Note that if this longer time constant were the dominant relaxation process at high power, two-level saturation would have occurred at a lower power in Fig.~\ref{F031806DGeq}(a).

The density-matrix model can also be used to predict the escape rates for the power series in Fig.~\ref{F031806DGtot}; calculations are plotted with solid lines in that figure.  The presence of the noise signal at $\wnm{0}{2}$ has little effect and inhomogeneous broadening decreases the escape rate at the lowest microwave power slightly.  The maximum value of $\Irf \paren{t}$ was again calculated for each power using the fit in Fig.~\ref{F031806Dfosc}.  The $T_\phi$ value of 16 ns used in the simulations was chosen to best reproduce the decay envelopes over the full range of powers.  At the highest power, there is a discrepancy in the oscillation maxima and minima, perhaps indicative of inaccurate parameters for levels \ket{3}\ and \ket{4}.  The envelope of the microwave turn-on is also reflected in the shape of the escape rate.

As a final test of the model, we examine multiphoton transitions in the system.  Figure~\ref{F010206H1}(c) shows a grayscale plot of the escape rate calculated with a seven-level density-matrix simulation.  Nearly all of the features seen in the data of Fig.~\ref{F010206H1}(a) are present in the simulation.  As the gray scales are identical, a small quantitative disagreement is visible, particularly for the \trans{0}{3}\ three-photon transition.  Figure~\ref{F010206H1}(d) shows the normalized power spectral density calculated from Fig.~\ref{F010206H1}(c); it agrees well with the data in Fig.~\ref{F010206H1}(b).

\begin{figure}
  \includegraphics[bb=0 0 366 540, clip, width=3.0in]{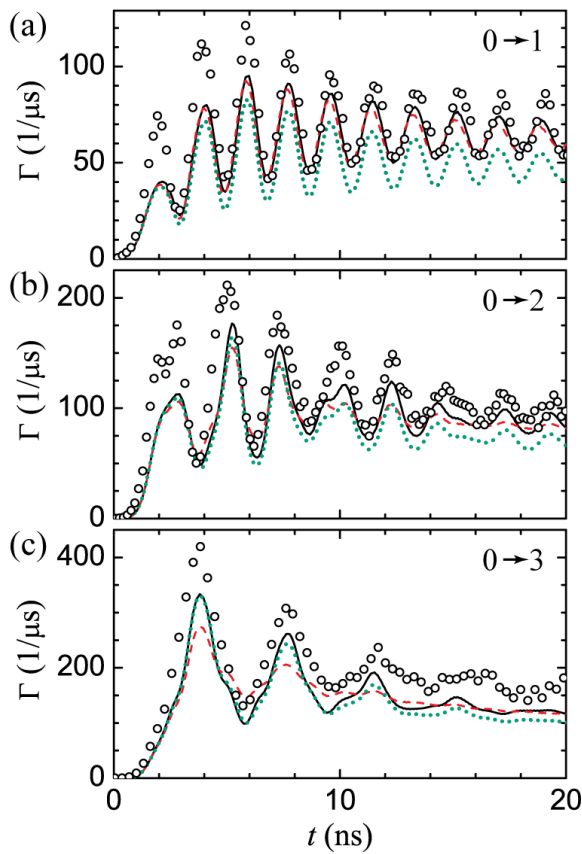}    \vspace{-0.0in}
  \caption{\label{F010206H1lines} (Color online) Comparison of possible damping scenarios for multi-photon transitions.  The measured escape rates from Fig.~\ref{F010206H1}(a) for the (a) \trans{0}{1}, (b) \trans{0}{2}, and (c) \trans{0}{3} transitions are plotted with circles.  The solid lines were calculated with the same simulation (with a common dephasing time $T_\phi$) whose results are shown in Fig.~\ref{F010206H1}(c).  The dotted lines were calculated for $T_1 = 10\ \ns$ and no additional dephasing, while the dashed lines correspond to $T_1 = 17\ \ns$ and pure dephasing with $T_\phi = 16\ \ns$ due to low frequency noise (see text).}
\end{figure}

Figure~\ref{F010206H1lines} shows line cuts of the time-domain data (circles) and simulation (solid lines) from Fig.~\ref{F010206H1} at bias currents of (a) 17.623, (b) 17.596, and (c) 17.571 \uA.  These values of $I_b$ correspond to the resonances of the one, two, and three-photon transitions at the high power at which the data were taken.  While the decay time and the long-time escape rate \Geq\ are reproduced well for the three transitions, the first few nanoseconds are not captured fully.  This could be due to distortion of the microwave current by the coupling capacitor \Crf\ or other line mismatches.

As with all of the simulations discussed in this section so far, the solid lines in Fig.~\ref{F010206H1lines} were calculated with $T_1 = 17\ \ns$ and $T_\phi = 16\ \ns$.  This gives $T_2 = 10.9\ \ns$ and a Rabi decay time roughly equal to the two-level value\cite{Torrey49a, Smith78a} of $T^\prime = \brac{1/\paren{2 T_1} + 1/\paren{2 T_2}}^{-1} = 13.3\ \ns$, consistent with measurements of the decay envelope of the escape rate.  If pure dephasing were not present then $T^\prime$ would be 22.7 ns, which is significantly longer than what is observed.  Although unlikely, our thermal measurement of $T_1 = 17\ \ns$ could be incorrect.  To examine this possibility, simulations with $T_1 = 10\ \ns$ and $T_2 = 2 T_1$ (which also give $T^\prime = 13.3\ \ns$) are shown with dotted lines in Fig.~\ref{F010206H1lines}.  They are nearly identical to the curves calculated with dephasing, although \Geq\ is somewhat smaller.  Thus this data set alone cannot rule out dissipation-limited decoherence.  However, the shorter $T_1$ reduces the prediction for \Geq\ by roughly 15\% over the full range of measured powers in Fig.\ \ref{F031806DGeq}, suggesting that additional dephasing is instead affecting the Rabi oscillations.

While dephasing is needed to faithfully reproduce features of the experimental measurements, its origin is unclear.  In the simulations, we assume that each off-diagonal term of the density matrix decays with a common dephasing time of $T_\phi = 16\ \ns$ (along with decoherence due to dissipation, which is transition dependent).  If this dephasing at high power were due to low frequency noise (similar to the effects of inhomogeneous broadening discussed before),\cite{Martinis03a, Xu05b} the corresponding dephasing operator $D$ in \eqa{h} could be expressed with $L = \sum \wnm{0}{n} / \wnm{0}{1} \ket{n} \bra{n}$ and $\lambda = 2 / T_\phi$ (similar to the harmonic oscillator number operator).  With this choice of operator $L$ and strength $\lambda$, the \trans{0}{1}\ dephasing is unchanged.  Simulations with this sort of damping are shown with dashed lines in Fig.~\ref{F010206H1lines}.  While the \trans{0}{1}\ Rabi oscillations are  modeled well, there is far too much decoherence for the higher order transitions.  This should also be true for other implementations of low frequency noise that have dephasing rates that scale as $(d \wnm{n}{m}/d I_b)^2$.  Thus having the ability to measure a wide range of transitions can reveal additional information about decoherence.

\section{Discussion and Summary}
\label{SSummary}

We have presented measurements taken on a dc SQUID operated in such a way that one of the junctions behaves much like a simple current-biased junction.  We find that the simplified Hamiltonian of \eqa{a}\ gives an accurate description of the qubit dynamics.  Our observation of subtle features of this model provides further confidence that it can be applied to design the gates needed for quantum computation.  The analysis presented here can also be extended to describe the behavior of the other types of superconducting qubits mentioned in Sec.~\ref{SIntro}.

We performed several checks on the model, including measuring the Rabi oscillation frequency for one and two-photon transitions and comparing the data with predictions based on the rotating wave Hamiltonian of \eqa{d}.  We find good agreement, despite the model not containing any damping.  Thus the resonance shifts in multilevel Rabi oscillation frequencies are mainly determined by the anharmonic level structure and not the $T_1$ and $T_2$ of the qubit.  These multilevel effects were identified under strong driving where it is clear that they could lead to errors in simple two-state rotations.  Future experiments could mitigate these effects by proper pulse shaping and operating at low microwave power.\cite{Steffen03a, Amin06a, Lucero08a}

We also used a density-matrix simulation, which included the effects of decoherence, to calculate the tunneling escape rate.  The results of the simulation agreed well with experimental Rabi oscillation data which were acquired with the device biased so that even the ground-state escape rate \Gn{0}\ was measurable.  While this simple measurement (which ends with the qubit in the finite voltage state on every repetition) is not an ideal projective measurement to the qubit basis states \ket{0}\ and \ket{1}, it is particularly well-suited for the purposes of the current work.  For one, it is extremely sensitive to excited state population due to a microwave drive or thermal transitions.  The leakage information contained in Fig.~\ref{F031806DGeq} would be considerably more difficult to obtain from the pulsed single-shot measurements\cite{Cottet02a, Claudon04a} that excel at determining the total population not in the ground state; nonetheless, these techniques have been successfully used to measure the second excited state population precisely.\cite{Palomaki07a, Claudon07a, Lucero08a}  With the simple tunneling measurement, the large ratio $\Gn{n+1} / \Gn{n}$ provides a natural way of distinguishing which excited state is populated.  In addition, as nothing is done to the qubit to initiate a measurement, this method is not limited by a measurement fidelity and does not suffer from the potential problems associated with changing the bias location in order to perform state readout.\cite{Cooper04a, Zhang06a, Palomaki07a, Claudon07a}

The main drawback to the escape rate measurement is that it does not directly produce individual state populations.  However, the density-matrix simulations do provide each level's contribution to the total escape rate and its occupation probability.  An example for a \trans{0}{1}\ oscillation is shown in Figs.~\ref{F032106FN10}(b) and \ref{F032106FN10}(c).  The plots suggest that for high power, the escape rate during a Rabi oscillation is dominated by the contributions $\rho_{22} \Gn{2}$ and $\rho_{33} \Gn{3}$ from states \ket{2}\ and \ket{3}.  The figure also shows that $\rho_{22} \paren{t}$ and $\rho_{33} \paren{t}$ have nearly the same form as $\rho_{11} \paren{t}$, albeit with a much smaller oscillation amplitude.\cite{Amin06a}  Thus changes in $\Gamma$ are reflective of the underlying oscillation of \ket{1}, which is why we believe the frequency analysis in Fig.~\ref{F031806Dfosc} is valid.

The populations in Fig.~\ref{F032106FN10}(c) have been normalized at each time $t$.  While the probability of the junction leaving the supercurrent state during a microwave pulse depends strongly on \Irf\ and the bias conditions, $\rho_{tot} = 0.42$ at $t = 30\ \ns$ for this data set.  Thus, while we wish to use tunneling as a probe of $\rho_{nn}$, it is clearly playing a large role in the evolution of the system.  However, repeating the density-matrix simulation in the absence of tunneling (\textit{i.e.}, $G \dm = 0$) results in a less than 5\% change in the normalized populations of \ket{0}\ and \ket{1}\ during the oscillation.  The subsequent decay is significantly different without the fast decays due to escape.  These insights from the simulation hinge on the assumption that tunneling does not affect $T_1$ and $T_\phi$.

A few remarks should be made about the many input parameters required for the simulations.  For example, the seven-level density-matrix calculation shown in Fig.~\ref{F010206H1}(c) needed seven escape rates, six energy-level spacings, and 28 matrix elements of $\gamma_1$, all of which are functions of $I_b$.  We determined these from the properties of a tilted washboard potential defined by $I_{01}$ and $C_1$, which were measured independently by low power spectroscopy.  To accurately reproduce the measured escape rate (particularly at low power), we also had to add several other features to the simulation, each motivated by separate measurements.  In particular, we added a microwave source at \wnm{0}{2}\ to mimic noise (whose magnitude was found from the microwave-free escape rate), inhomogeneous broadening due to bias noise (estimated from spectroscopic resonance widths), and a finite time resolution (consistent with the bandwidth of the detection electronics).  The dissipation time $T_1 = 17\ \ns$ was estimated by an independent thermal escape rate experiment,\cite{Dutta04a} while the dephasing time $T_\phi = 16\ \ns$ (and the way it was incorporated into the simulation) was chosen to maximize the agreement with the Rabi data.  For simplicity, both time constants were assumed to be independent of frequency.  Removing this condition in the analysis could yield additional information about the decoherence in the system, as the oscillations are most sensitive to noise at the Rabi frequency.\cite{Martinis03a, Ithier05a}  Finally, the conversion between the microwave power at its generator and the current through the qubit junction was itself calibrated by the observed oscillation frequency, and the resulting factor was allowed to be a function of microwave frequency.  Thus a wide range of multilevel phenomena was explained with just a few truly free parameters.

In addition, we found that the output of the simulations (whether that be an oscillation frequency or escape rate) converged as the number of levels $N$ was increased.  For any of the five or seven-level simulations discussed here, an additional level produced a change too small to be detected by the experiment; more levels were required at high power or low $I_b$ for the solution to converge satisfactorily.  Surprisingly, the highest included levels are predicted to lie above the barrier.  However, small errors in the energy levels or matrix elements for the highest levels do not produce large changes in the final escape rates.

Certain features of the data were not reproduced by the simulations.  Although not discussed here, the transition spectra of our Nb qubits show a series of splittings (all less than 10 MHz wide), similar to but smaller than those reported in other superconducting qubits.\cite{Simmonds04a, Plourde05a, Claudon07a, Dong07a}  The extra degrees of freedom responsible for these features were not included in the device Hamiltonian, except perhaps in some effective way through $T_1$ and $T_\phi$ of the density-matrix simulation.  For weak coupling and high microwave power, the impact of individual two-level systems on Rabi oscillations is expected to be small.\cite{Meier05a, Ashhab06a}  However, it is unclear if a bath of quantum systems could be responsible for the anomalous decay in Fig.~\ref{F032106FN10}.  As other groups have observed strong dissipation in Nb trilayer junctions,\cite{Martinis02a, Lisenfeld07a} it is likely that the wiring insulation or other details of the fabrication are playing a critical role.\cite{Martinis05a}  While further work is needed to identify (and potentially eliminate) the actual microscopic sources of decoherence in this system, the multilevel features in the dynamics of our qubit are largely understood.

We have benefited greatly from discussions with B.~K.~Cooper, P.~R.~Johnson, H.~Kwon, J.~Matthews, K.~D.~Osborn, B.~S.~Palmer, A.~J.~Przybysz, R.~C.~Ramos, C.~P.~Vlahacos, and H.~Xu.  The work was funded by the National Science Foundation through the QuBIC Program, the National Security Agency, and the state of Maryland through the Center for Nanophysics and Advanced Materials, formerly the Center for Superconductivity Research.

\providecommand{\BIBYu}{Yu}


\begin{thebibliography}{66}
\bibitem[{NIS()}]{NIST}
\bibinfo{note}{Certain commercial equipment, instruments, or materials are
  identified in this paper to specify the experimental procedure adequately.}
\end{thebibliography}
\end{document}